\documentclass[10pt,a4paper]{article}
\usepackage[left=18mm, top=10mm, right=14mm, bottom=20mm, nohead, foot=10mm]{geometry}
\usepackage[affil-it]{authblk}

\let\OLDthebibliography\thebibliography
\renewcommand\thebibliography[1]{
  \OLDthebibliography{#1}
  \small
  \setlength{\parskip}{0pt}
  \setlength{\itemsep}{0pt plus 0.3ex}
}

\usepackage{multicol}

\usepackage{caption}

\usepackage{misccorr}

\usepackage[english]{babel}
\usepackage{amsmath,amssymb}
\usepackage{textcomp}
\usepackage{ulem}

\usepackage{xcolor}

\usepackage{hyperref}

\hypersetup{colorlinks=true, linkcolor=[rgb]{0.6 0 0}, citecolor=[rgb]{0 0.4 0}, urlcolor=[rgb]{0 0 0.6}}

\usepackage{graphicx}
\usepackage{epsfig}

\newenvironment{Figure}
  {\par\medskip\noindent\minipage{\linewidth}\captionsetup{type=figure}}
  {\endminipage\par\medskip}
\usepackage{epstopdf}

\definecolor{ao(english)}{rgb}{0.0, 0.5, 0.0}

\title{\large \textbf{
Chiral Magnetic Skyrmions with Arbitrary Topological Charge 
\\(\textit{Skyrmionic Sacks}) 
}}
\author[1]{\normalsize Filipp~N.~Rybakov\thanks{f.n.rybakov@gmail.com}}
\author[2]{\normalsize Nikolai~S.~Kiselev}

\affil[1]{\small
Department of Physics, KTH-Royal Institute of Technology, SE-10691 Stockholm, Sweden}

\affil[2]{\small
Peter Gr\"{u}nberg Institut and Institute for Advanced Simulation, Forschungszentrum J\"{u}lich and JARA, D-52425 J\"{u}lich, Germany}

\date{}

\begin{document}

\maketitle
\vspace{-3.0\baselineskip}

\renewcommand{\abstractname}{\vspace{-\baselineskip}}
\begin{abstract}
We show that continuous and spin-lattice models of chiral ferro- and antiferromagnets provide the existence of an infinite number of stable soliton solutions of any integer topological charge.
A detailed description of the morphology of new skyrmions and the corresponding energy dependencies are provided. 
The considered model is general, and is expected to predict a plethora of particle-like states which may occur in various chiral magnets including atomic layers, e.g., PdFe/Ir(111), rhombohedral GaV$_4$S$_8$ semiconductor, B20-type alloys as Mn$_{1-x}$Fe$_x$Ge,  Mn$_{1-x}$Fe$_x$Si, Fe$_{1-x}$Co$_x$Si, Cu$_2$OSeO$_3$, acentric tetragonal Heusler compounds. 
\end{abstract}

\begin{multicols}{2}[]

The existence of stable localized magnetic vortices in the presence of so-called chiral interactions was predicted in 1989~\cite{Bogdanov_89}.
Nowadays, it is common to use the term \textit{chiral magnetic skyrmions} for these vortices.
In contrast to baby skyrmions~\cite{Piette,Weidig}, and superconducting skyrmions~\cite{Garaud_drag}, chiral magnetic skyrmions are not related to the model of baryons proposed by T.\,H.\,R.~Skyrme~\cite{Skyrme_1,Skyrme_2} neither explicitly nor implicitly~\cite{Babaev_2012}. 
Similarly to the aforementioned skyrmions, these magnetic vortices can be characterized by a non-zero topological charge $Q$.
The corresponding micromagnetic Hamiltonian contains competing terms of different powers of spatial derivatives with respect to the order parameter mimicing the Skyrme mechanism of stabilization~\cite{Bogdanov_1995}. 
The presence of competing terms highlights the model of a chiral magnet among many other models in nonlinear physics where the Hobart-Derrick theorem forbids the existence of stable localized solutions~\cite{Rajaraman}.

The research in this field has gotten a powerful impetus after the direct observation of magnetic skyrmions in cubic chiral magnets of B20-type employing Transmission Electron Microscopy~\cite{Yu_10}.
Later, the existence of magnetic skyrmions has been confirmed in many other materials~\cite{Romming_13,Tokunaga_15,Kez_15,Nayak_17,Kanazawa2017}. 
Several new phenomena have been reported recently: different approaches for the nucleation of magnetic skyrmions~\cite{Romming_13,Jin_17}, electric current induced motion of skyrmions~\cite{Yu_skyrmion_flow}, attractive and repulsive inter-skyrmion interactions~\cite{Loudon_2018,Du_2018}. 
The possible utilization of magnetic skyrmions in spintronic devices is also under intensive study~\cite{Kiselev11,Fert_13,Zhang_15(1),Muller,Zheng_17,Bessarab_sk_lifetime}.

A fundamental question concerning the diversity of possible skyrmion solutions remains unaddressed in the literature despite the large number of experimental and theoretical works published within the last decade. 
Indeed, some of two-dimensional (2D) models such as: baby Skyrme~\cite{Piette,Weidig} and isotropic ferromagnet~\cite{BP} possess a wide variety of solutions with an arbitrary integer $Q$. However, in the frame of underlying model of chiral magnet, it was assumed that the diversity of skyrmions is limited to topological charge $Q\!=\!\pm 1$~\cite{Nagaosa_review, SOC_review, Fert_review}. 
This underlying model~\cite{Bar,Bak,Bogdanov_89} contains only three energy terms: the exchange interaction, chiral Dzyaloshinskii-Moriya interaction (DMI)~\cite{Dzyaloshinskii,Moriya} and potential term. 
The latter one represents the interaction with an external magnetic field $\mathbf{B}_\mathrm{ext}$ and/or magnetic anisotropy, meaning the presence of energetically preferable directions for magnetization vector, $\mathbf{M}(\mathbf{r})$.

First, the static skyrmion solution with $Q\!=\!-1$ has been reported in Ref.~\cite{Bogdanov_89}.
There is a generalization of this solutions nowadays known as Bogdanov-Hubert $k\pi$-vortices~\cite{Bogdanov_99} characterized by alternating $Q\!=\!-1$ and $Q\!=\!0$ for odd and even $k$, respectively.
For convenience, we employ the term $\pi$-\textit{skyrmion}~\cite{Hagemeister} referring to $1\pi$-vortex with $Q\!=\!-1$ and the term \textit{skyrmionium}~\cite{Finazzi,Komineas} for $2\pi$-vortex with $Q\!=\!0$.

Following the classification of skyrmion solutions given in Ref.~\cite{Bogdanov_89} one can show (see also the discussion in Ref.~\cite{Hoffmann}) that by applying trivial operations of reflection and rotation to the magnetization vectors, the solutions corresponding to the different type of crystal symmetries can be always mutually transformed one into another. 
Thus, these solutions belong to the same class.  
An important consequence of such classification is that stable magnetic textures recently discovered in tetragonal Heusler alloy~\cite{Nayak_17} and named ``antiskyrmion'' in fact belongs to the same class of $\pi$-skyrmion solutions.
Moreover, recently W.~Koshibae and N.~Nagaosa suggested~\cite{Koshibae_Nagaosa} that in a conventional model of a chiral magnet the coexistence of stable skyrmion solutions with $Q\!=\!-1$ and $Q\!=\!1$ is impossible.

In this work it is shown that in fact, a conventional model of chiral magnet possesses an infinite number of skyrmions with different value and sign of topological charge and diverse morphology.

\begin{figure*}
\centering
\includegraphics[width=17cm]{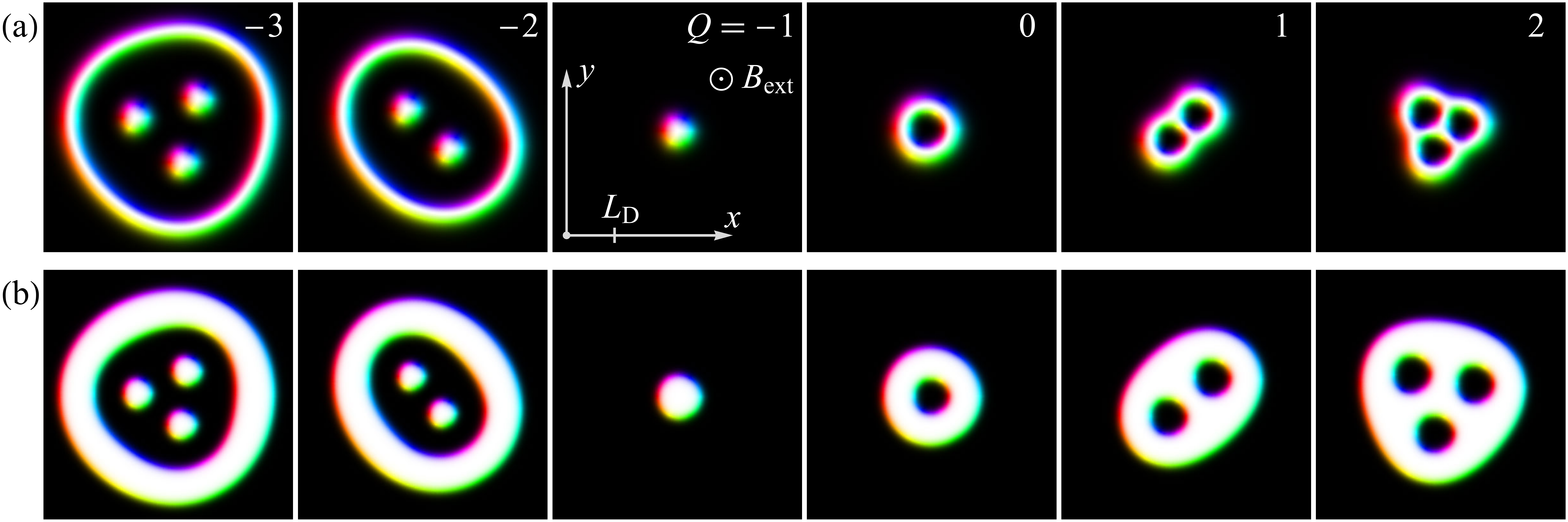}
\caption{\small
Morphology of stable chiral skyrmions with topological charges $Q\!=\!-3,-2,...,2$. 
Top row of images (a) corresponds to zero magnetocrystalline anisotropy ($u\!=\!0$) in external magnetic field applied perpendicular to the plane, $h\!=\!0.65$; Bottom row of images (b) corresponds to the case of uniaxial anisotropy, $u\!=\!1.3$ and zero external field, $h\!=\!0$. 
All images are given in the same scale. 
Colors encode the direction of the $\mathbf{n}$-vectors according to standard scheme~\cite{StSp}: black and white denote up and down spins respectively, red-green-blue reflect azimuthal angle with respect to $x$-axis. 
}
\label{Fig_zoo}
\end{figure*}

The energy functional for the continuous two-dimensional model of a chiral magnet can be written in the following form~\cite{Bogdanov_89}: 
\begin{equation}
E =\! \int\! \Big(  
\mathcal{A}\sum_{i}\left( \mathbf{\nabla} n_i \right)^2 + \mathcal{D}\,w(\mathbf{n}) + U(n_z) \Big)\,t\,\mathrm{d}x\,\mathrm{d}y,
\label{Ham}
\end{equation} 
where  $\mathbf{n}\!\equiv\!\mathbf{n}(\mathbf{r})$ is a continuous unit
vector field ($n_x^2+n_y^2+n_z^2\!=\!1$) determining the direction of the magnetization, $\mathbf{n}\!=\!\mathbf{M}/M_\mathrm{s}$, $\mathcal{A}$ and $\mathcal{D}$ are the micromagnetic constants for exchange and DMI, respectively. $t$  represents the layer thickness.  
The last term in~(\ref{Ham}) is the potential term, which consists of the Zeeman energy: $U_\mathrm{Z}\!=\!B_\mathrm{ext}\,M_\mathrm{s}\,(1 - n_z)$, and/or the energy density of uniaxial anisotropy $U_\mathrm{a}\!=\!K\,(1 - n_z^2)$.
The DMI term $w(\mathbf{n})$ represents a linear combination of Lifshitz invariants:
$$\Lambda_{ij}^{(k)} = n_i\frac{\partial n_j}{\partial r_k} - n_j\frac{\partial n_i}{\partial r_k},$$
which in turn is defined by the underlying symmetry of the crystal.

The results presented below are valid for a very wide class of chiral magnets of different lattice symmetries:
for the systems with N\'{e}el-type modulations~\cite{Romming_13,Kez_15,Romming_15} where 
$w(\mathbf{n}) = \Lambda_{xz}^{(x)} + \Lambda_{yz}^{(y)}$, for the tetragonal compounds of D$_{2\mathrm{d}}$ symmetry~\cite{Nayak_17} with 
$w(\mathbf{n}) = \Lambda_{zy}^{(x)} + \Lambda_{zx}^{(y)}$ and for the crystals with bulk-type DMI and Bloch-type modulations~\cite{Yu_10,Yu_11,Yu_15} where 
$w(\mathbf{n})  = \Lambda_{zy}^{(x)} + \Lambda_{xz}^{(y)} + \Lambda_{yx}^{(z)} = \mathbf{n}\cdot\left(\mathbf{\nabla}\times\mathbf{n}\right)$.
For systems with bulk-type DMI, e.g., B20-type crystals, the solutions discussed below should be considered as a new kind of skyrmion tubes (or strings) that were previously known only for the case of a $\pi$-skyrmion~\cite{Rybakov_13,Milde,Rybakov_15,Leonov_16,Kagawa,Du_2018}.
Note, when the term $\Lambda_{yx}^{(z)}$ play a significant role~\cite{Rybakov_13,StSp}, one should solve a three-dimensional problem.

For the case of magnetic multilayers and heterostructures with interface-induced DMI~\cite{Wanjun_review}, the demonstration of the stability of the solutions displayed further might suggest that isolated $\pi$-domains (also known as bubble-skyrmions) ~\cite{Wanjun,Woo,Boulle,McVitie} belong to the category of chiral skyrmions. 
In the eventuality where such a scenario fails, one can conclude that the magnetic dipole-dipole interaction represents the core mechanism for the stabilization of those domains. 
This in turn indicates that such objects are magnetic bubbles~\cite{bubbles_book} rather than chiral skyrmions.

For convenience, all parameters such as the lengths, the value of the external magnetic field and the energy are given in relative dimensionless units of:
the equilibrium period of helical spin spiral~\cite{helix,Bogdanov_11}, $L_\mathrm{D}\!=\!4\pi\mathcal{A}/|\mathcal{D}|$, the critical field of the cone spiral transition into saturated state~\cite{Bogdanov_11}, $B_\mathrm{D}\!=\!\mathcal{D}^2/(2M_\mathrm{s}\mathcal{A})$ and the energy $E_0=2\mathcal{A}t$. 
Thereby, only two dimensionless parameters completely define the state of the system:
$$h=B_\mathrm{ext}/B_\mathrm{D}, \quad u=K/(M_\mathrm{s}B_\mathrm{D}).$$

The localized solutions are the excitations on the homogeneous background, in other words $\mathbf{n}(\mathbf{r})\rightarrow\mathbf{n}_0$ for ${|\mathbf{r}|\rightarrow\infty}$. 
Thus, the domain of definition of the order parameter $\mathbf{n}(\mathbf{r})$ can be mapped to a sphere which can be associated with a Riemann sphere ($\mathbb{R}^2\!\cup\!\{\infty\}\!\leftrightarrow\!\mathbb{S}^2$). 
The space of the order parameter $\mathbf{n}$ is in turn a sphere $\mathbb{S}_{\mathrm{spin}}^2$.
The map $\mathbb{S}^2\!\rightarrow\!\mathbb{S}_{\mathrm{spin}}^2$ leads to an homotopy classification of localized solutions in 2D with topological invariants related to an integer index
\begin{equation}
Q = \frac{1}{4\pi}\int \Big( \mathbf{n}\cdot \left(\partial_x\mathbf{n}\times\partial_y\mathbf{n}\right)  \Big)\mathrm{d}x\mathrm{d}y.
\label{Q}
\end{equation}
For topologically non-trivial textures $Q\!\neq\!0$ a smooth transformation into a homogeneous state $\mathbf{n}(\mathbf{r})\!=\!\mathbf{n}_0$ is impossible. 
We follow the sign convention for a topological charge given in Ref.~\cite{Melcher_14}, and for definiteness we assume the polarity with $\mathbf{n}_0\!=\!(0,0,1)$. 

Note, equation (\ref{Q}) often represents a useful quantity for the estimation of the number of $\pi$-skyrmions in the clusters appearing due to the geometrical confinement of the sample~\cite{Leonov_target,Zhao_16}, under the pressure of surrounding non-uniform helical phase~\cite{Muller_PRL}, under the impact of spatially varying external stimuli~\cite{Wenjing}, different pinning effects~\cite{Koshibae_disordered}, etc. 
The topological charge $|Q|$  of these clusters can be greater than one. 
However, it is essential to distinguish between a cluster of particles and a single skyrmion with topological charge $Q$. 
This aspect can be easily explained by analogy to the Skyrme model for atomic nuclei. 
In this model the topological charge, $B$ corresponds to the baryon number of the nuclei. 
For instance, the helium-4 atom has $B\!=\!4$ and its nucleus is a single skyrmion, while deuterium molecule D$_2$ has the same $B\!=\!4$ but includes two nuclei and thus represents the cluster of two skyrmions.

The stable solution for $\pi$-skyrmion with the energy below~\cite{Bogdanov_89} and above~\cite{Ivanov} the 
energy of the saturated state was found earlier.  
Such $\pi$-skyrmion represents only a single example of the vast variety of topologically non-trivial solutions, 
see Fig.~\ref{Fig_zoo}. 
For the case of $u\!=\!0$ and high magnetic fields $h\!\geq\!2$ it was shown~\cite{Melcher_14} that $\pi$-skyrmion is 
energetically most favorable among all other hypothetical configurations with $Q\!\neq\!0$. 
It will be shown below that $\pi$-skyrmion and skyrmionium with $Q\!=\!0$ are the key elements or ``building blocks'' for the whole variety of other skyrmions.

One may highlight two main reasons why stable skyrmion solutions for $Q\!<\!-1$ and $Q\!>\!0$ have been overlooked earlier: (i) the inter-particle repulsion of $\pi$-skyrmions~\cite{Bogdanov_1995} and (ii) an extremely limited set of axisymmetric critical points~\cite{Li} of the Hamiltonian~(\ref{Ham}). 
Both effects obstruct the merging of $\pi$-skyrmions into single particle, and instead result in them moving apart which is so-called dichotomy~\cite{Melcher_IFF}. 
Consequently, naive attempts to obtain stable $Q\!=\!-N$ skyrmion in numerical simulation by using $N$ isolated $\pi$-skyrmions as initial guess cannot lead to the nucleation of a large single skyrmion.

To find energetically stable solutions, we performed a direct energy minimization of the functional~(\ref{Ham}) based on a nonlinear conjugate gradient method implemented for NVIDIA CUDA architecture. 
We used a finite-difference discretization scheme of the fourth-order with a meshes density varying from $640^2$ to $5120^2$ nodes~[Section~\ref{Accuracy}].
The values of the variables $\mathcal{A}$ and $\mathcal{D}$ have been chosen such that the parameter $L_\mathrm{D}$ equals 52 inter-node distances.

To obtain a $Q\!<\!\!-1$ skyrmion, we put $N_\mathrm{cores}\!=\!|Q|$  number of $\pi$-skyrmions inside a ``sack'' representing a closed $2\pi$ domain wall, \text{i.e} skyrmionium which has topological charge, $Q\!=\!0$ (see Fig.~\ref{Fig_zoo}). 
Supplementary movies~[Section~\ref{Video}] illustrate the crafting of the initial states for different anticipated morphologies of skyrmion solutions and their energy minimization.
The closed domain wall plays the role of the shell of the skyrmion and has a tendency to shrink down to the equilibrium size of skyrmionium. 
Inter-particle repulsion of $\pi$-skyrmions in turn prevent such shrinking.
Similar to the effect of surface tension, the balance of external and internal pressures results in the stability of this spin texture.
For skyrmion with $Q\!>\!0$ the role of a ``sack'' or a shell is played by a closed $\pi$ domain 
wall which possesses a nonzero topological charge $Q\!=\!-1$ as a $\pi$-skyrmion.  
The domain within the closed loop has magnetization opposite to the surrounding ferromagnetic background. 
Due to the opposite polarity, each $\pi$-vortex inside such ``sack'' has self topological charge equal to one. 
As a result, the total topological charge $Q\!=\!(N_\mathrm{cores}-1)$, where the amount of cores is equal to the number of ``holes''. 
In Fig.~\ref{Fig_zoo}, for $Q\!=\!1$ and $2$, these cores look as ``holes'' inside the white domain. 
We found solutions with absolute values of $Q$ equal to units, tens, hundreds and even thousands~[Section~\ref{Big}]. 
Thereby, there is every reason to believe that $Q$ could be equal to any arbitrary large integer number.

The dependence of the skyrmion energy as a function on its topological charge is found to be well approximated 
by a piecewise linear function for small $|Q|$, while some points slightly deviate from the linear law 
[Figs.~\ref{Fig_is} and~\ref{Fig_as}]. 
The same linear law dependence $E(Q)$ is known to be a good approximation in the baby Skyrme model~\cite{Piette,Weidig}, 
while for an isotropic ferromagnet model~\cite{BP} the relation is strictly linear.
Our analysis~[Section~\ref{Big}] shows that the curves $E_\textrm{aspt} = E_0\,(\alpha_{(\pm)}N_\mathrm{cores} + \beta_{(\pm)}\sqrt{N_\mathrm{cores}})$ are good candidates for the true asymptotics when $Q\!\rightarrow\!\pm\infty$.
A detailed numerical analysis with a high precision confirm the equation $\alpha_{(-)}=E_{Q=-1}/E_0$ (for details see Section~\ref{Big}).

\begin{Figure}
\centering
\includegraphics[width=\columnwidth]{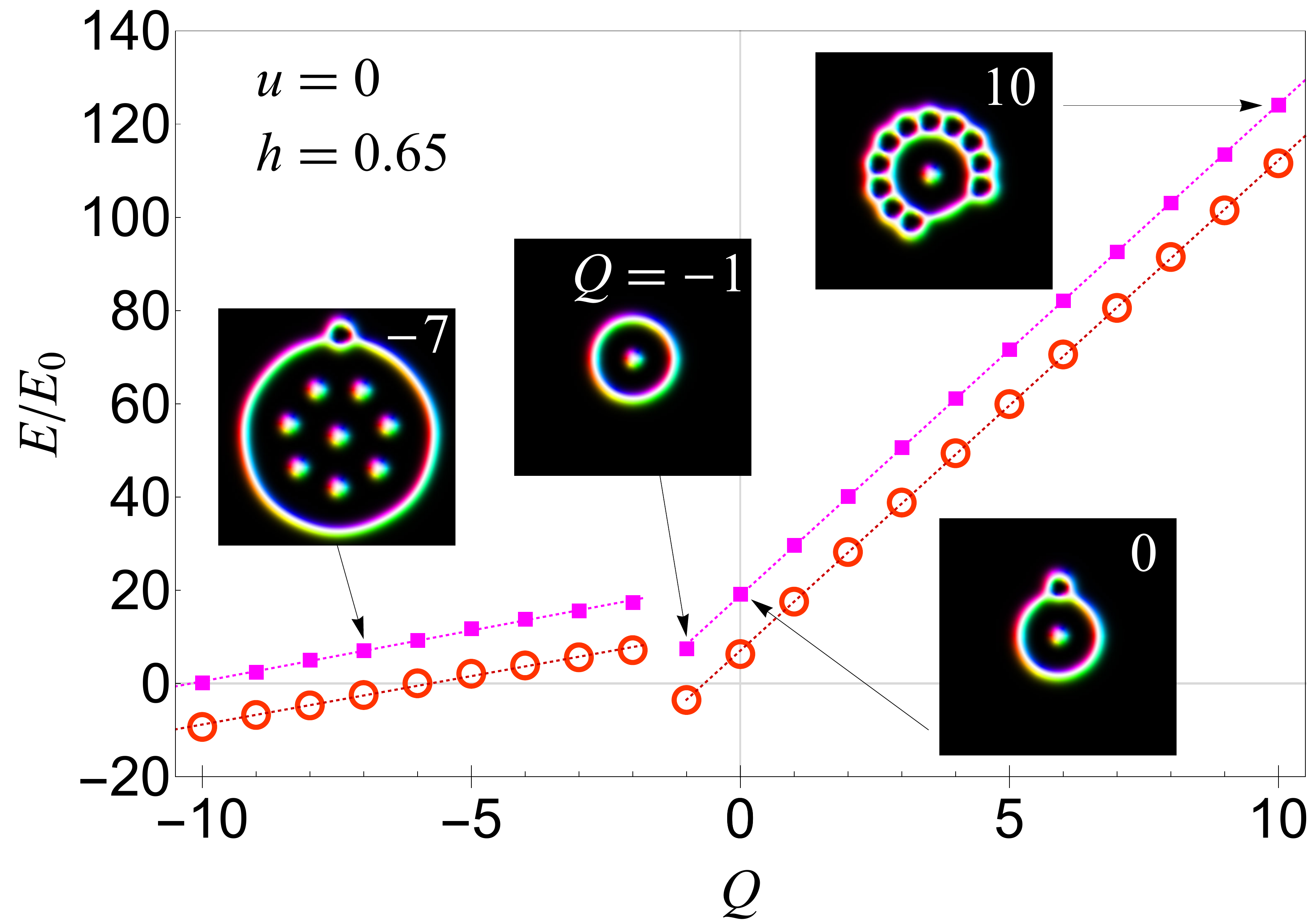}
\caption{\small
The energy of skyrmions, $E$ as function of topological charge, $Q$ for the case of a magnet without magnetocrystalline anisotropy, $u\!=\!0$. 
Open circles are the lowest energy solutions for each particular $Q$, solid squares -- solutions with higher energies but nearest to the lowest energy state. 
The dotted lines are linear fits for corresponding sets of points. 
}
\label{Fig_is}
\end{Figure}

Let us first consider the case of zero magnetic anisotropy, $u\!=\!0$ [Fig.~\ref{Fig_is}]. 
The value of external field $h$ has been chosen to be above the field of the elliptical instability~\cite{Bogdanov_1994} and below the field of the thermodynamic stability of $\pi$-skyrmion, $E_{Q=-1}\!<\!0$~\cite{Bogdanov_1994JMMM}. 
The right branch of the ``spectrum'' for $Q\!\geq\!-1$ increases monotonically with $Q$. 
In contrast to that the left branch of the ``spectrum'' ($Q\!<\!-1$) displays the opposite behavior and the energy decreases with $|Q|$. 
It is because the global energy minimum in this case corresponds to a hexagonal  lattice of $\pi$-skyrmions~\cite{Bogdanov_1994JMMM} and skyrmions with $Q\!\ll\!-1$, on the left branch of ``spectrum'' form a kind of lattices inside their shells~[Section~\ref{Big}].

It worth to mention that there are also the solutions with higher energies. 
In Fig.~\ref{Fig_is}, we have shown only the energies of those states which correspond to the smallest energy shift. 
Some of those solutions are shown in inset. 
Note, $Q\!=\!-1$ skyrmion shown in inset corresponds to earlier known solution of $3\pi$-vortex~\cite{Bogdanov_99}. 
We believe that solutions with lowest energies (open circles in Fig.~\ref{Fig_is}) represent true minimizers in the corresponding topological sectors.

For the case of a nonzero out-of-plane uniaxial anisotropy, Fig.~\ref{Fig_as}, we use the value $u\!=\!0.65$ which in accordance to Ref.~\cite{Romming_15} corresponds to the bilayer of PdFe on an Ir(111) single crystal substrate. 
The calculations were performed for two values of $h$ for which $E_{Q=-1}\!>\!0$. 
Both branches of the spectrum now demonstrate the same trend~[Fig.~\ref{Fig_as}]. 
Above certain values of $h$, the energy of some points on the left branch of the spectrum become higher than the critical Dirichlet energy~\cite{BP,Tretiakov} which is shifted on the corresponding energy of skyrmionium, see solid line in Fig.~\ref{Fig_as}(a),(b). 
This indicates that such solutions loss their stability, see for instance the point marked with four arrows in Fig.~\ref{Fig_as}(b). 
Such instability can be explained as follows. 
The pressure from the shell becomes too high and leads to the shrinking~\cite{Tretiakov} of the internal $\pi$-skyrmions. 
This may result in blow-up behaviour of the solutions~\cite{Galaktionov} with an increasing magnetic field. 
From that one may conclude that in the case of such a critical phenomena, a certain solution of higher energy branch becomes the minimizer for corresponding $Q$, see inset in Fig.~\ref{Fig_as}(b).

\begin{Figure}
\centering
\includegraphics[width=1.0\columnwidth]{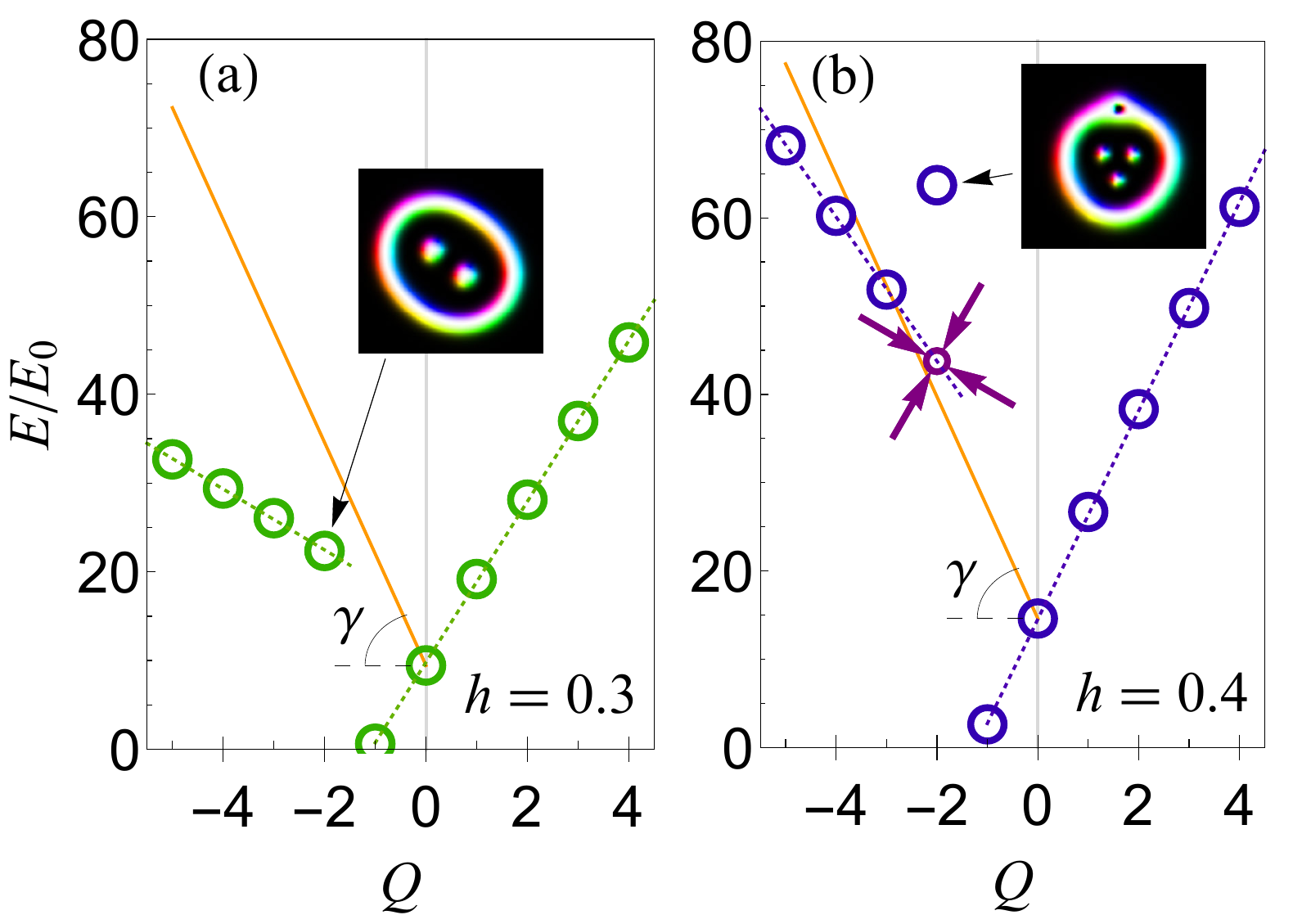}
\caption{\small
The energy of skyrmions, $E$ as function of topological charge, $Q$ for the case of uniaxial 
anisotropy, $u\!=\!0.65$ in external magnetic field $h\!=\!0.3$ (a) and $h\!=\!0.4$ (b).
The slope of the solid line is defined as $\tan(\gamma)\!=\!4\pi$. 
The point marked by four thick arrows in (b) corresponds to the solution expected from the linear fit of the energy dependence and with morphology shown in inset in (a), but precise calculations shows instability of expected skyrmion. 
The morphology of stable skyrmion with $Q\!=\!-2$ is shown in inset.
}
\label{Fig_as}
\end{Figure}

The above discussed solutions are not restricted to the continuum model and can be generalized to spin-lattice models of chiral ferro- and antiferromagnet. 
In the Section~\ref{Lattice}, we present solutions for such models, and discuss their morphology and key features.

In conclusion, we have shown that the standard micromagnetic model with chiral Dzyaloshinskii-Moriya interaction allows the existence of chiral magnetic skyrmions with any integer topological charge. 
The morphology of the new solutions with $Q\!<\!-1$ and $Q\!>\!0$ are described in details, and are found to be sufficiently different from the systems where the inter-particle attraction naturally leads to the formation of clusters~\cite{Garaud_drag,Garaud_CP2,Rozsa,Kharkov}. 
The energies of skyrmions with a high topological charge are found to be comparable to each other and controllable by an external magnetic field. 
The latter suggests that the direct observation of a large variety of new particle-like states presented here should be accessible in experiment. 
We suppose that the nucleation of new skyrmions can possibly be realized with a Scanning Tunneling Microscope equipped with the magnetic tip of special shape~\cite{Wieser} or by means of a time-dependent external magnetic field or a current-induced spin torque effect in geometrically confined systems. 
The stabilized solitons can be considered as information bit carriers in a skyrmion racetrack memory and may extend a newly proposed concept of ``two-particles'' for binary data encoding, as skyrmion-bobber chains in cubic chiral magnets~\cite{Zheng_17} or sequences of skyrmions and antiskyrmions in spatially anisotropic ultra thin films~\cite{Hoffmann}.

Finally, one has to emphasize that the found solutions for $|Q|\!>\!1$ are natural high charge skyrmions for corresponding model, while $Q\!=\!1$ solution presented in this work should be considered as actual antiparticle to $\pi$-skyrmion.

After the first version of this work~\cite{first_version}, D.~Foster et al. reported an independent work~\cite{Foster} where authors report on similar theoretical findings and provide an experimental evidence of so-called ``skyrmion bags'' observed in liquid crystals which mimics skyrmions with high topological charge addressed in this work.

\end{multicols}

\setcounter{equation}{0}
\setcounter{figure}{0}
\setcounter{table}{0}

\renewcommand{\thesection}{\normalsize\text{A}\arabic{section}}
\renewcommand\theequation{\text{S}\arabic{equation}} 
\renewcommand\thefigure{\text{S}\arabic{figure}}   
\renewcommand{\thetable}{\text{S}\arabic{table}}

\part*{\centering \large Appendices}\label{app}

\section{\normalsize Accuracy of finite difference schemes}\label{Accuracy}

\begin{figure}[ht]
\centering
\includegraphics[width=0.5\textwidth]{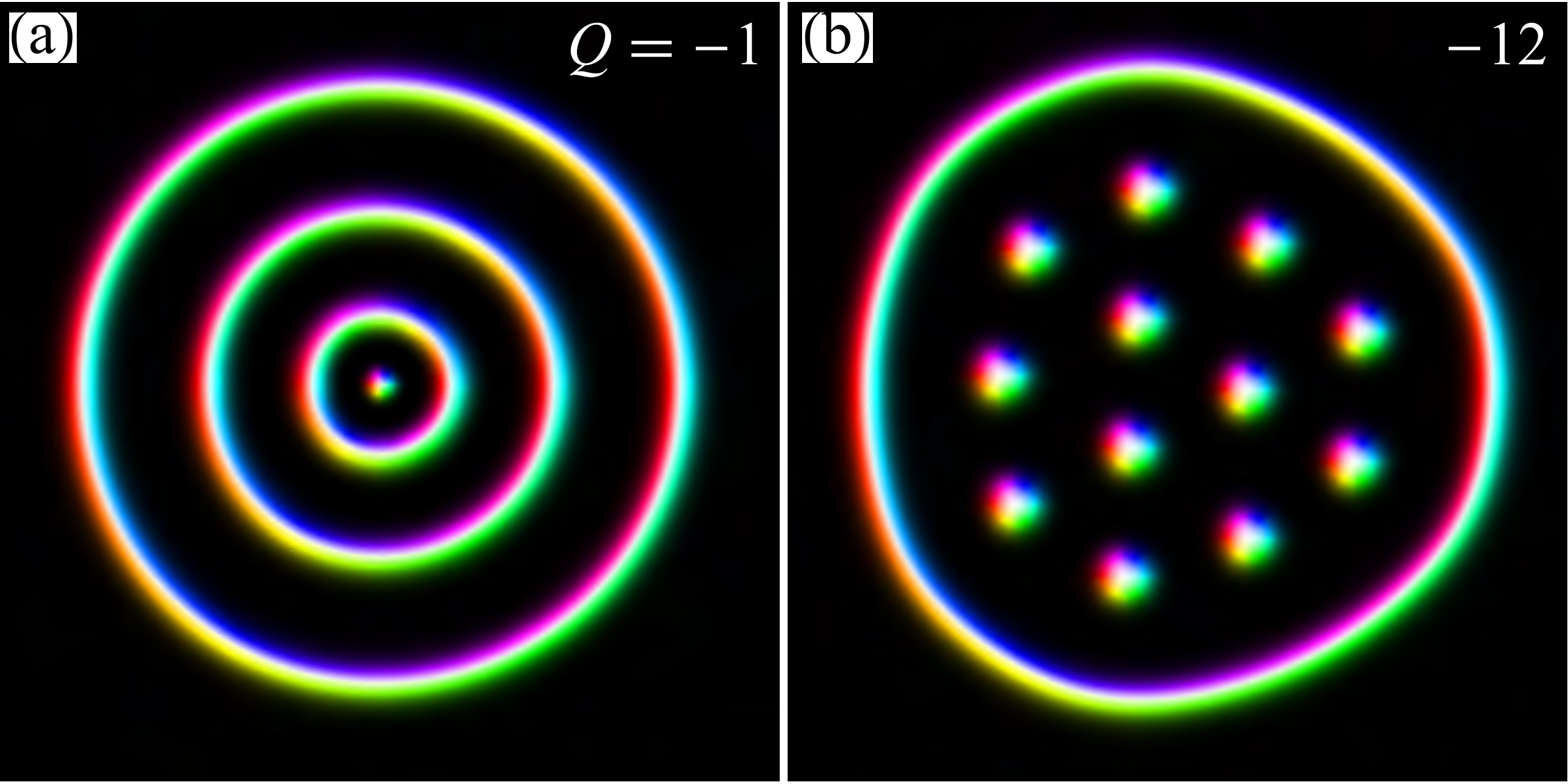}
\caption{\small
Example of two solitons of nearly identical size but different morphology and topological charge: axisymmetric $7\pi$-vortex~\cite{Bogdanov_99} with $Q\!=\!-1$ (a) and skyrmion with $Q\!=\!-12$ (b). 
Both solutions corresponds to the case $u\!=\!0$, $h\!=\!0.65$. 
Both images are given in the same scale. 
The color code is identical to that one used in Figs.~1-3 in the main text.
} 
\label{sup_Fig_7pi}
\end{figure}

In this section, we discuss the accuracy of the numerical methods employed to calculate various skyrmionic solutions with different topological charge and morphology presented in this work.
We compared the results of energy minimization obtained by our method with the second-order finite difference schemes implemented in most open-source  software for micromagnetic simulations such as MuMax3~\cite{MuMax3}. 
A high accuracy numerical scheme used in our work is essential for the study of a number of aspects: the stability of the solutions close to blow-up, energy of skyrmions with extremely big topological charge, etc. 
Moreover, we provide additional calculations with a very high accuracy reducing the relative error in energy down to 10$^{-6}$.
These calculations can be taken as benchmarks and compared with the outputs provided by other methods.
Such high accuracy can be achieved only for axisymmetric solutions~\cite{Bogdanov_99}, where the problem can be reduced to ordinary differential equation.
An example of axisymmetric solution is depicted in Fig.~\ref{sup_Fig_7pi}(a). 
Furthermore, we show a non-axisymmetric solution of a comparable size with a more complex morphology in Fig.~\ref{sup_Fig_7pi}(b).
The results obtained with our precise method for axisymmetric solutions can be taken as a reference to define a threshold for more general solutions with lower symmetries.

The Hamiltonian (1) in the main text can be rewritten in dimensionless units:
\begin{equation}
\mathcal{E} = \frac{E}{E_0}=\int \Big(  
\frac{1}{2}\sum_{i}\left( \mathbf{\nabla} n_i \right)^2 + 2\pi\,w(\mathbf{n}) + 4\pi^2\,u\, (1-{n_z}^2) + 4\pi^2\,h\,(1-n_z)  \Big)\mathrm{d}\mathtt{x}\mathrm{d}\mathtt{y},
\label{Ham_sup}
\end{equation} 
where $\mathtt{x}=x/L_\mathrm{D}$, $\mathtt{y}=y/L_\mathrm{D}$,  $E_{0}$ and $L_\text{D}$  are defined in the main text.

In the case of axisymmetric solitons~\cite{Bogdanov_89,Bogdanov_99}, the solution of the problem~(\ref{Ham_sup}) can be reduced to a second order nonlinear non-autonomous ordinary differential equation:
\begin{equation}
\underbrace{
\frac{d^2\theta}{d\rho^2} 
+ \frac{1}{\rho}\frac{d\theta}{d\rho} 
- \frac{1}{\rho^2}\sin(\theta)\cos(\theta)
}_\mathrm{exchange}
+
\underbrace{
 \frac{4\pi}{\rho}\sin(\theta)^2
}_\mathrm{DMI}
-
\underbrace{
4 \pi^2 u \sin(2\theta)
}_\mathrm{uniax.\,anis.}
-
\underbrace{
4 \pi^2 h \sin(\theta)
}_\mathrm{Zeeman}
= 0,
\label{DE_sup}
\end{equation}
where $\theta$ is the polar angle of magnetization vector, i.e. $n_z = \cos(\theta)$ and $\rho$ is the radial coordinate. 

The soliton solutions of equation~(\ref{DE_sup}) are exponentially localized~\cite{BKY}, even in absence of the DMI contribution~\cite{Kovalev}.
The true asymptotic of such solutions behaves as a Macdonald function~\cite{Voronov,Leonov_NJP}:
\begin{equation}
\theta(\rho) \sim \frac{1}{\sqrt{\rho}} \exp\left( -2\pi\sqrt{2u + h}\,\rho \right) 
\quad\mathrm{for}\,\,\,\rho\rightarrow\infty.
\label{asymp}
\end{equation}
This exponential decay of the solution renders the error introduced by the finite-size domain simulations negligible. 
The discretization scheme plays a major role for the achievement of the required high accuracy. 
We are particularly interested in the behaviour of the error with respect to the size and morphology of the skyrmion texture. 
A higher accuracy can be achieved for one-dimensional (1D) problem~(\ref{DE_sup}) since our solution $\theta$ depends only on $\rho$.
Contrary to 2D case the solution of such 1D problem does not require a considerable computational efforts, and high accuracy results can be obtained on very dense meshes with inter-node distance $\sim\!0.001$ (about a 1000 nodes per $L_\mathrm{D}$).
Such high-accuracy solutions then can be used as benchmarks to verify the accuracy of other methods.

The equation~(\ref{DE_sup}) can be solved numerically by means of explicit integration relying of the Runge-Kutta method~\cite{Bogdanov_99}. 
Assuming $\theta(0)\!=\!k\pi$, where $k$ is an integer, the proper value of the parameter $\theta^\prime(0)\!=\!d\theta/d\rho |_{\rho=0}$ can be found by shooting method.
Thus, the solution is uniquely ``encoded'' in a single number $\theta^\prime$ and every overshooting/undershooting should lead to a distortions of the $\theta(\rho)$ profile which is expected to decrease monotonically to zero as $\rho\!\rightarrow\!\infty$.
This property of the explicit integration method can be used to verify the correctness of the results obtained with other methods. 
In particular, we found the solution of 1D problem~(\ref{DE_sup}) by \textbf{unconstrained} Nonlinear Conjugate Gradient (NCG) minimization method for corresponding Hamiltonian:
\begin{equation}
\mathcal{E}_\mathrm{1D} = 2\pi\int\displaylimits_0^\infty 
\left(
\frac{1}{2}\left(\frac{d\theta}{d\rho}\right)^2 + 
\frac{1}{2\rho^2}\sin(\theta)^2 + 
2\pi\frac{d\theta}{d\rho} + 
\frac{\pi}{\rho}\sin(2\theta) +
4\pi^2 u \sin(\theta)^2 + 
4\pi^2 h (1 - \cos(\theta)) 
\right)\rho\,\mathrm{d}\rho.
\end{equation}
Very large simulation domain was used, $0\!\leq\!\rho\!\leq\!10$ and very small inter-node distance $\Delta\rho\!=\!0.005$. 
The values at the boundaries, $\theta(0)$ and $\theta(10)$ was fixed to $k\pi$ and $0$ respectively.
A finite-difference scheme of the fourth order of accuracy was designed assuming that $\theta_i$ and its spatial derivative, $\theta^\prime_i$ at each node, $i$ are independent variables. 
According to our estimates the relative error in the calculation of the energy marked in bold font shown in the Table~\ref{tab1} does not exceed $10^{-6}$. 
For additional verification, we used the value $\theta^\prime_{i=0}$ (the value in the first node for which $\rho\!=\!0$) from the found solution as an initial input value $\theta^\prime(0)$ for the integration with fourth-order Runge-Kutta method (RK4).
We used the integration step $\Delta\rho\!=\!10^{-7}$ and found during further shooting procedure that at least the first six significant digits of $\theta^\prime_{i=0}$ are accurate.
All calculations for 1D problem were carried out in double-precision format (64 bits) for floating-point operations.

The contributions of exchange and Dzyaloshinskii-Moriya interaction terms at each $(i,j)-$th node with coordinates $(\mathtt{x},\mathtt{y})\!=\!(i\,\Delta{s},j\,\Delta{s})$ in 2D mesh are approximated using the values of unit vector field in eight neighboring nodes with 
$(\mathtt{x}\!\pm\!\Delta{s},\mathtt{y}\!\pm\!\Delta{s})$ 
and 
$(\mathtt{x}\!\pm\!2\Delta{s},\mathtt{y}\!\pm\!2\Delta{s})$, 
where $\Delta{s}$ -- inter-node distance. 
The corresponding energy contributions represent the products of the $\mathbf{n}$-vector projections at each node and its eight neighbor nodes multiplied by specific factors, see for instance Ref.~\cite{Buhrandt13} and Supplementary Materials in Ref.~\cite{Milde}.  
For testing purposes, we have also implemented in our code the conventional second-order finite difference scheme. 
For direct energy minimization, we used \textbf{constrained} NCG algorithm where the constraint $\mathbf{n}^2\!=\!1$ is naturally satisfied because of using the atlas for the manifold corresponding to the space of the order parameter.
The manifold itself represents a two-dimensional sphere $\mathbb{S}_{\mathrm{spin}}^2$ while the atlas is composed of two coordinate charts each of which corresponds to stereographic projection from one of two poles of the sphere.
Here, we refer to this advanced numerical scheme as ``Atlas''.
Conceptually such scheme is similar to the idea of describing the macrospin in the frame of stereographic projections with the ability to switch between projections from the both poles, presented in Ref.~\cite{Horley}.
The key feature of the ``Atlas'' scheme is that each individual spin is defined in one of two coordinate chart independently on other spins.
A more detailed description of the method and criteria for switching between charts for individual spins can be found in Supplementary Materials of Ref.~\cite{Rybakov_15}.
Note, the most of the floating-point operations in our code have been implemented in single-precision format (32 bits). 
This allows to reach high performance on GPU.

In Table~\ref{tab1}, we present the comparison of the results obtained with MuMax3 where 2-nd order finite-difference scheme is implemented~\cite{MuMax3} (the script is provided in an ancillary file \href{https://arxiv.org/src/1806.00782v2/anc/mumax3-script.mx3}{\underline{mumax3-script.mx3}}), different implementation of Atlas method with 2-nd order (see ``Atlas\,$\backslash$\,2'' columns with $\Delta{s}\!=\!1/52$ and more dense mesh with $\Delta{s}\!=\!1/104$), 4-th order finite-difference scheme (see ``Atlas\,$\backslash$\,4'' column), and one-dimensional approach (see ``1D\,$\backslash$\,4'' column).
It is seen that the method chosen in current work provides the best accuracy with an error lower than $0.04\%$ as for  $Q\!=\!-1$ skyrmions as for more complex textures.
In contrast, the 2-nd order finite-difference scheme widely used in micromagnetic software shows significant error of about $8\%$ for the same relatively dense mesh.
Increasing twice the mesh density in each of the dimensions reduces this error only by factor four as expected for the second-order scheme.
A simple estimate suggests that one requires to increase the mesh density fourteen times more in each dimension to provide an accuracy comparable to our fourth-order scheme. 
Finally, it worth mentioning that the calculations for the textures with characteristic size of $\sim\!10L_\mathrm{D}$ (see for instance Fig.~\ref{sup_Fig_7pi}) with second-order discretization scheme on the meshes with $\Delta{s}\!\sim\!0.01L_\mathrm{D}$ provides an absolute error in energy calculations higher than value of $E_0$. 
Therefore, for quantitative analysis of such and larger textures the second-order discretization scheme becomes unreliable.

\begin{table}
\centering
\caption{\small 
The energies of several solitons calculated by different methods.} \label{tab1}
\resizebox{\columnwidth}{!}{%
\begin{tabular}{ |c|r|c|c|c|c|c|c|c|c| } 
\hline
\multicolumn{4}{|c}{} & 
\multicolumn{5}{|c|}{Energy, $\mathcal{E}$} & 
\multicolumn{1}{c|}{}  \\ 
\hline
texture type & 
$Q$ &
$u$ & 
$h$ & 
MuMax3\,$\backslash$\,2 & 
Atlas\,$\backslash$\,2 &
Atlas\,$\backslash$\,2 &
Atlas\,$\backslash$\,4 & 
1D\,$\backslash$\,4 &
$\theta^\prime(0)$ \\ 
 & & & &  
 $\Delta{s}=1/52$ & 
 $1/52$ &
 $1/104$ &
 $1/52$ & 
 $1/200$ &
  \\
\hline
axisymmetric skyrmion & -1 & 0 & 0.65       & -3.522 & -3.527 & -3.556 & -3.565  & \textbf{-3.56497} & -4.553561 \\
\hline
$2\pi$-vortex (skyrmionium) & 0 & 0 & 0.65       & 6.476 & 6.472 & 6.322 & 6.274  & \textbf{6.27244} & -2.424450  \\
\hline
$7\pi$-vortex & -1 & 0 & 0.65       & 55.45 & 55.46 & 53.63 & 53.02 & \textbf{53.0071} & -8.847806 \\
\hline
skyrmion & +1 & 0 & 0.65      & 17.85  & 17.79 & 17.59 & 17.52 &  not applicable & not applicable\\
\hline
skyrmion & -3 & 0 & 0.65       & 6.466 & 6.462 & 5.808 & 5.595 &  not applicable & not applicable \\
\hline
skyrmion & -12 & 0 & 0.65       & -12.54 & -12.54 & -13.91 & -14.37 &  not applicable & not applicable \\
\hline
axisymmetric skyrmion & -1 & 0.65 & 0.3     & 0.673 & 0.673 & 0.6341 & 0.6220 &  \textbf{0.621763} & -3.012659  \\
\hline
axisymmetric skyrmion & -1 & 0.65 & 0.4    & 2.666  & 2.666 & 2.628 & 2.617 &  \textbf{2.61621} & -4.561866  \\
\hline
\end{tabular}
}
\end{table}


\section{\normalsize Real-time simulations}\label{Video}

One of the key features implemented in our code is the graphic user interface with an interactive regime allowing the \textit{in situ} control of the magnetic configurations as well as an easy way to construct a large variety of initial states. 
In particular, when being in the interactive regime, one can flip  the spins inside a certain area under the mouse pointer. 
This option provides an efficient approach for construction of complex initial configurations composed of domains with a magnetization pointed either up or down. 
After a certain number of iterations of the energy minimization routine, the initial configuration converges to one of the nearest energy minimum. 
Beside the calculation of standard termination criteria~\cite{Gill_textbook} one can also perform an \textit{in situ} examination 
of the stability by introducing small excitations and perturbations to the simulated spin texture.

In order to emphasize the isomorphism of systems with different Lifshitz invariants, we prepared three distinct movies illustrating the case of 
C$_{\mathrm{nv}}$ (\href{http://www.youtube.com/watch?v=LOiDfXhGalw}{\underline{movie~1}}), 
D$_{2\mathrm{d}}$ (\href{http://www.youtube.com/watch?v=qo75nEE0N7Q}{\underline{movie~2}}) and 
D$_{\mathrm{n}}$ (\href{http://www.youtube.com/watch?v=Nf2Nd7KduAk}{\underline{movie~3}}) symmetries.


\section{\normalsize Big and extremely big skyrmions}\label{Big}

\begin{figure}[h]
\centering
\includegraphics[width=1.0\textwidth]{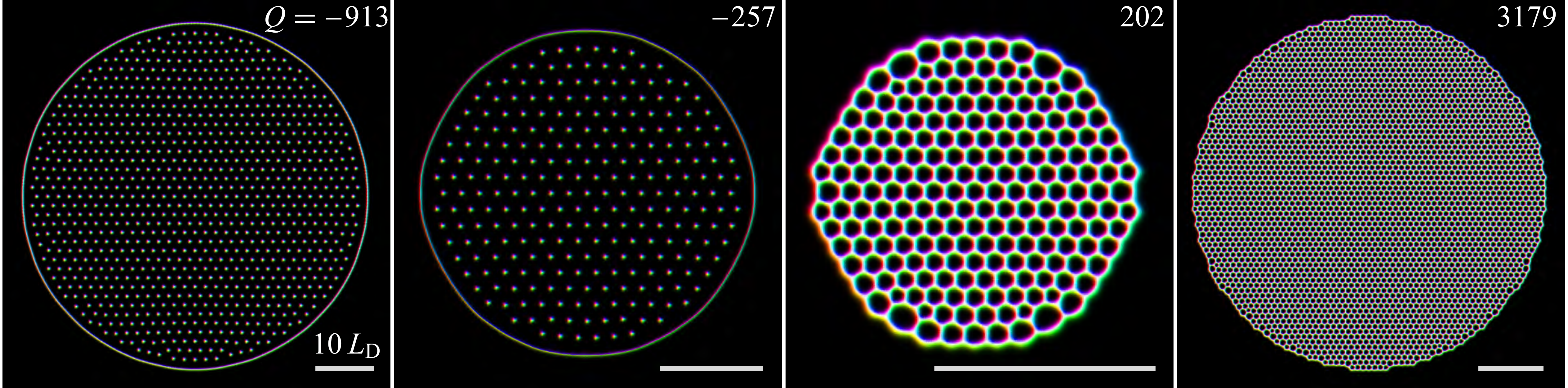}
\caption{\small
Morphology of stable chiral skyrmions with high topological charges in case of magnetic field applied perpendicular to the plane, $h\!=\!0.65$ and zero magnetocrystalline anisotropy, $u\!=\!0$. 
Note, the scale is different for all figures. 
} 
\label{sup_Fig_zoo}
\end{figure}

It the case of $|Q|\!\gg\!1$, the kernel of the skyrmion -- its major internal part consists of tightly packed cores representing $\pi$-vortices. 
The shell of such heavy skyrmions which represents a $\pi$- or $2\pi$ domain wall for positive and negative $Q$, respectively, occupies relatively small area along the outer perimeter, see Fig.~\ref{sup_Fig_zoo}.
When increasing of the number of cores $N_\textrm{cores}$ the structure of the skyrmion kernel becomes more regular while the area engaged by the kernel increases proportionally to $N_\textrm{cores}$. 
As a result, the energy of the skyrmion kernel tends to be proportional to $N_\textrm{cores}$ while  contribution from the boundary is proportional to the perimeter of the skyrmion and is proportional to $\sqrt{N_\textrm{cores}}$.
Thereby, the asymptotic behaviour of the energy of the skyrmions with increasing $|Q|$ should have the following form:
\begin{equation}
\frac{E_\textrm{aspt}}{E_0}=
\begin{cases}
\alpha_{(-)}|Q|   + \beta_{(-)}\sqrt{|Q|} \quad &(Q\ll{-1}),\\
\alpha_{(+)}(Q+1) + \beta_{(+)}\sqrt{Q+1} \quad &(Q\gg{1}),
\end{cases} 
\label{aspt}
\end{equation}
where $\alpha_{(\pm)}$, $\beta_{(\pm)}$ are the constants which depend only on $u$ and $h$.

\begin{figure}[h]
\centering
\includegraphics[width=1.0\textwidth]{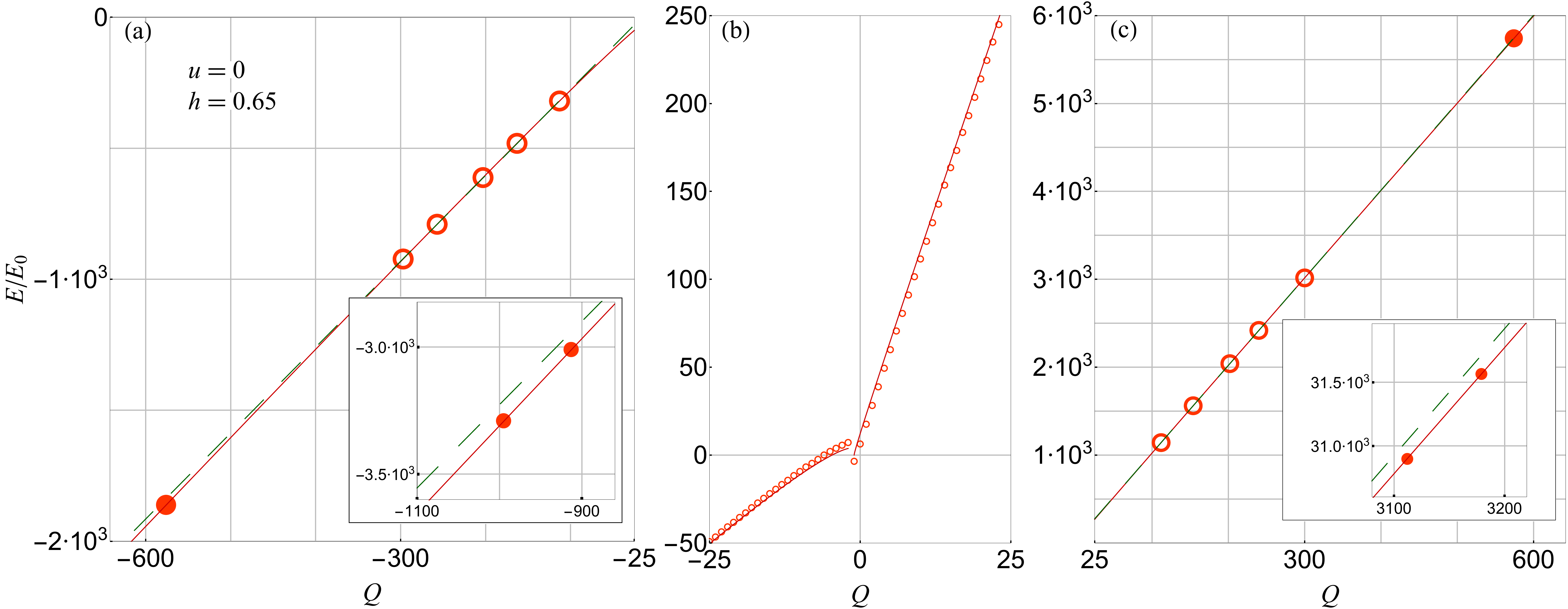}
\caption{\small
The energy of skyrmions, $E$ as function of topological charge $Q$ for: $Q\!\ll\!-1$ (a), $-25\!\leq\!Q\!\leq\!25$ (b) and $Q\!\gg\!1$ (c). 
The solid curves are fit by~(\ref{aspt}) for the points marked with empty circles in (a) and (c). 
The dashed lines in (a) and (c) are the linear fit for the same points.
The solid circles in (a), (c) and in corresponding insets were not taken in to account in fitting process and are shown to illustrate a high quality of the fit obtained with Eq.~(\ref{aspt}) and deviation of $E(Q)_{|Q|\!\gg\!1}$ from the assumption of linear dependence.
} 
\label{sup_Fig_E_on_Q}
\end{figure}

For the careful verification of~(\ref{aspt}), we first calculated ten skyrmions (five -- for negative $Q$ and five -- for positive $Q$) with relatively high topological charges in the range $100\!\leq\!|Q|\!\leq\!300$ (see empty circles in Figs.~\ref{sup_Fig_E_on_Q}(a) and \ref{sup_Fig_E_on_Q}(c)). 
Then assuming that such values of $|Q|$ are sufficiently large for the energy to be slightly different from the asymptote fitted with~(\ref{aspt}) and obtain the following fitting parameters: $\alpha_{(-)}=-3.552$, $\beta_{(-)}=7.663$, $\alpha_{(+)}=9.885$,  $\beta_{(+)}=2.308$. 
The dependencies corresponding to~(\ref{aspt}) are represented as solid curves in  Figs.~\ref{sup_Fig_E_on_Q}(a)-\ref{sup_Fig_E_on_Q}(c). 
Finally, in order to verify the expected asymptotic behavior, we have calculated the energies corresponding to the skyrmions with extremely high $|Q|$. 
As seen from Figs.~\ref{sup_Fig_E_on_Q}(a) and \ref{sup_Fig_E_on_Q}(c), the agreement is excellent.

To emphasize the deviation of $E(Q)$ from the linear dependence, we plotted results of the linear fit with the same points assuming $E\!\approx\!c_1|Q|\!+\!c_2$, see the dashed lines in Figs.~\ref{sup_Fig_E_on_Q}(a) and \ref{sup_Fig_E_on_Q}(c). 

Despite the fact that for large $|Q|$, the cores form a triangular lattice, the corresponding unit cell is different from the one of the skyrmion lattice phase also known as a \textit{skyrmion crystal}. 
In particular, the inter-skyrmion distances for an equilibrium skyrmion lattice is different from the inter-cores distance found in the kernels of big skyrmions.
In case of equilibrium skyrmion lattice, the particles are packed in such a way that the average energy density is minimized, while the number of particles is assumed to be unlimited in an infinite space. In contrast, the packing in the kernel of big skyrmion minimizes the total energy for a fixed number of cores inside a limited size domain.

In case of negative $Q$, if $|Q|$ increases then the pressure inside the sack decreases together with the curvature of the shell. 
Thereby, the stress of internal lattice should tends to zero as $Q\!\rightarrow\!-\infty$. For such a limiting case this lattice can be regarded in a first approximation as a set of individual non-interacting $Q\!=\!-1$ skyrmions,
which means that $\alpha_{(-)}$ is equal to $E_{Q=-1}/E_0$.
Our calculation gives $E_{Q=-1}/E_0\!=\!-3.565$ [Table~\ref{tab1}].
The corresponding discrepancy is only $0.4\%$ mostly due to the fact that the coefficients for the asymptote are obtained for a finite value of $|Q|$. 
For the case of uniaxial anisotropy ($u\!=\!0.65$, $h\!=\!0.3$) following the same procedure we found $\alpha_{(-)}=0.627$.
The corresponding energy $E_{Q=-1}/E_0\!=\!0.622$ [Table~\ref{tab1}].


\section{\normalsize Skyrmions in lattice models}\label{Lattice}

\subsection{\normalsize Chiral ferromagnet}

\begin{figure}[ht]
\minipage{0.45\textwidth}
\centering
\includegraphics[width=1.0\textwidth]{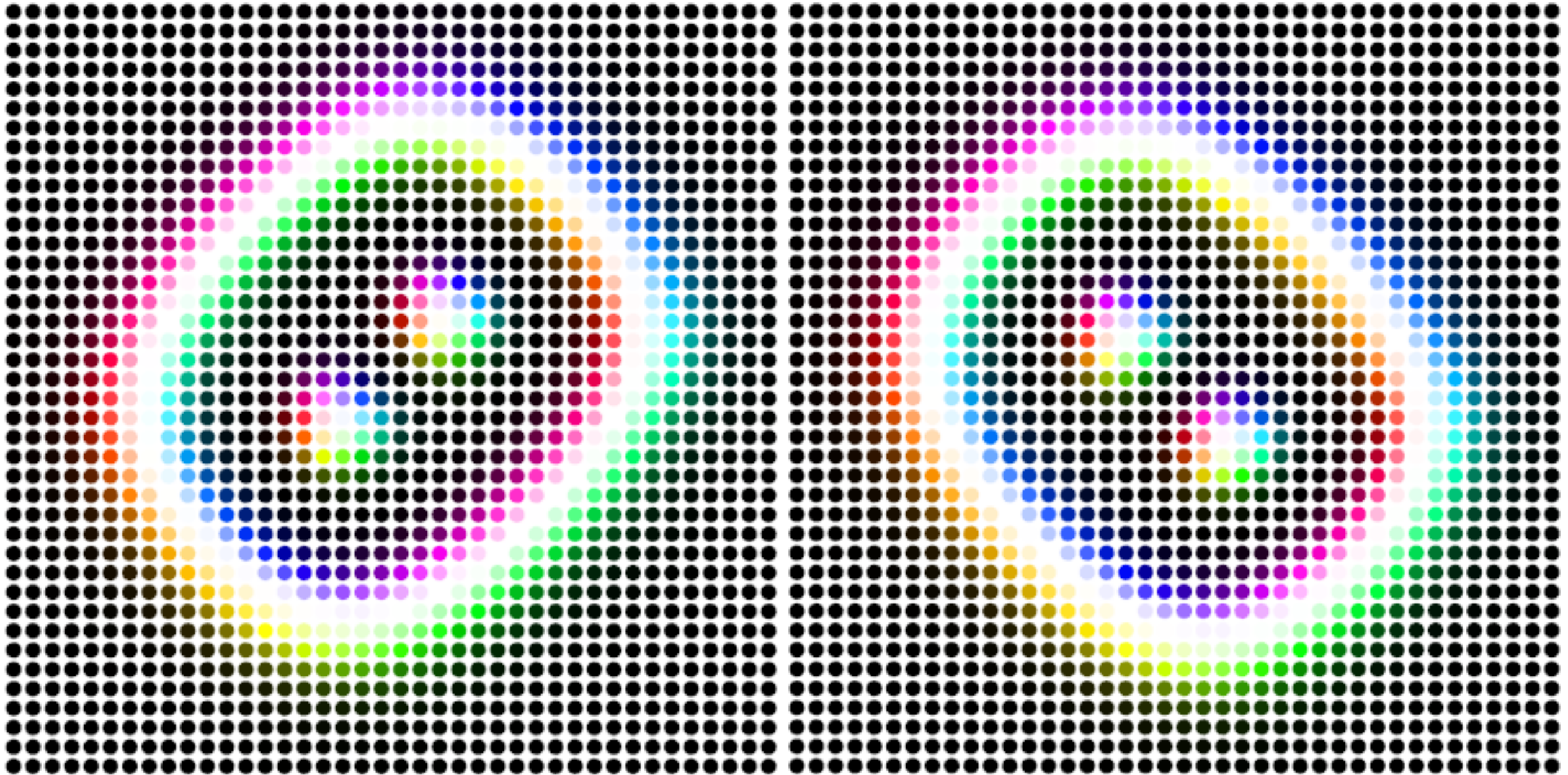}
\caption{\small
Two energetically equivalent states for skyrmion with $Q\!=\!-2$ on square lattice, for the case of zero magnetocrystalline anisotropy, $K_u\!=\!0$, $\mu_\mathrm{s}B_\mathrm{ext}\!=\!0.25$.
} 
\label{sup_Fig_D1}
\endminipage\hfill
\minipage{0.45\textwidth}
\centering
\includegraphics[width=1.0\textwidth]{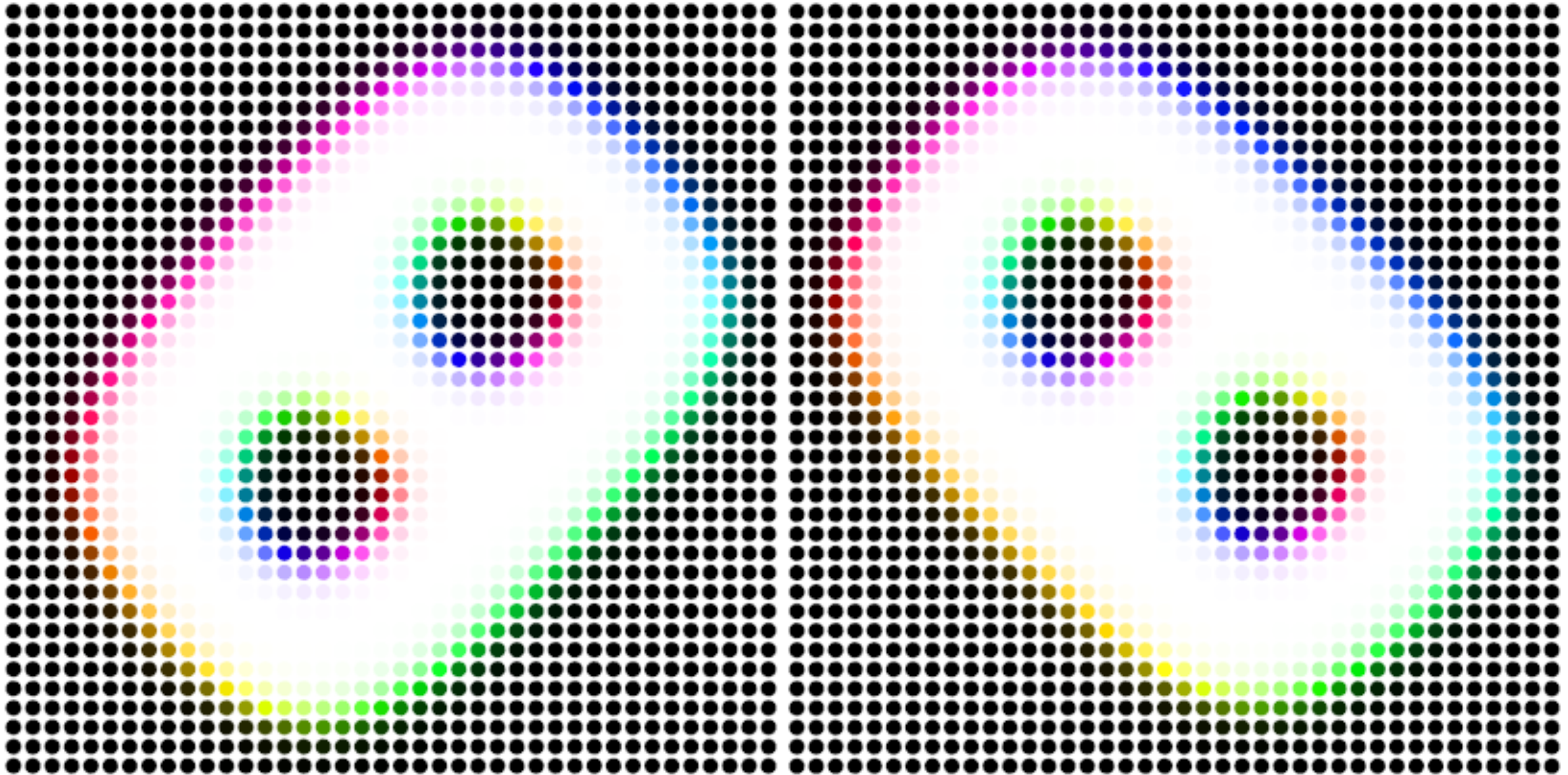}
\caption{\small
Two energetically equivalent states for skyrmion with $Q\!=\!1$ on square lattice for the case of no external field, $K_u\!=\!0.45$, $\mu_\mathrm{s}B_\mathrm{ext}\!=\!0$.
} 
\label{sup_Fig_D2}
\endminipage
\end{figure}

The results presented in this work which employs a high accuracy method for the quantitative analysis of continuous solutions remain valid in the discrete limit of classical spins on lattice. 
In addition to that, our results are also validated by the discrete approach for systems where the continuum approach (1) is unsuitable.
For illustration, we consider a standard spin lattice model of a chiral magnet~\cite{Sergienko,Han}:
\begin{equation}
\mathcal{H}\!=  - J \sum_{\left\langle ij\right\rangle, i>j }
 \mathbf{n}_i  \cdot  \mathbf{n}_j - \sum_{\left\langle ij\right\rangle, i>j } 
\mathbf{D}_{ij}  \cdot  [\mathbf{n}_i\! \times\!  \mathbf{n}_j]  
-  K_\mathrm{u} \sum_{i} {n}^2_{i,\mathrm{z}}
- \mu_\mathrm{s} \mathbf{B}_\mathrm{ext}\!\sum_{i}\mathbf{n}_i,
\label{Ham_latt}
\end{equation}
where $J$ is the exchange coupling constant, $\mu_\mathrm{s}$ -- magnetic moment of each spin. 
The unit vector $ \mathbf{n}_i$ defines the orientation of the spin at site $i$. 
The notation $\left\langle ij\right\rangle$, $i\!>\!j$ denotes that the summation runs over each nearest-neighbour pair once.
We assumed that each Dzyaloshinskii-Moriya pseudo vector $\mathbf{D}_{ij}$ is perpendicular to the bond between sites $i$ and $j$ and lies in the $(xy)$-plane. 
The modulus of vector $D\!=\!|\mathbf{D}_{ij}|$ is assumed to be fixed for all interacting pairs of spins.
For definiteness, we consider the case of 2D square lattice with lattice constant $a$; however, results presented below remain valid for other lattice symmetries as well.
The dominant interaction in the system is the ferromagnetic exchange, $0\!<\!D\!<\!J$.

In the absence of uniaxial anisotropy and external magnetic field  ($K_\mathrm{u}\!=\!0$, $B_\mathrm{ext}\!=\!0$) the ground state for~(\ref{Ham_latt}) is a spin spiral with a period $L\!=\!2\pi{a}/\mathrm{arctan}(D/J)$~\cite{Han}. 
For $J\!\gg\!D$ (and therefore $L\!\gg\!a$) continuous limit (1) can be considered as a valid approximation for the lattice  Hamiltonian~(\ref{Ham_latt}) with $\mathcal{A}\!=\!J/(2a)$, $\mathcal{D}\!=\!D/a^2$. The corresponding helix period $L_\mathrm{D}\!=\!2\pi{a}J/D$.
However, for $J\!\gtrsim\!D$ the continuum approach (1) representing a second order Taylor expansion of the lattice model~(\ref{Ham_latt}) becomes invalid.
For example, for $D\!=\!0.6J$ the period $L_\mathrm{D}$ turns to be underestimated by about $10\%$.
Thus, for such ratios between $J$ and $D$ the lattice effects
are relevant.

For our simulations, we used $J\!=\!1.0$, $D\!=\!0.6$.
An important feature of skyrmions in the lattice model is the discrete degeneracy of the solutions meaning that some in-plane directions are more preferable for texture alignment. 
In Figures~\ref{sup_Fig_D1} and~\ref{sup_Fig_D2}, we illustrate two possible skyrmion configurations for $Q\!=\!-2$ and $Q\!=\!1$, respectively.
The degree of degeneracy depends on the symmetry of the crystal lattice, and on the morphology of the skyrmion spin texture.
Note, the continuum model (1) is spatially isotropic and the energy of skyrmions does not depend on the orientation of the texture. For the calculation of topological charge on a discrete lattice we used the approach suggested in Ref.~\cite{Berg}.

\subsection{\normalsize Chiral antiferromagnet}

\begin{figure}[h]
\centering
\includegraphics[width=1.0\textwidth]{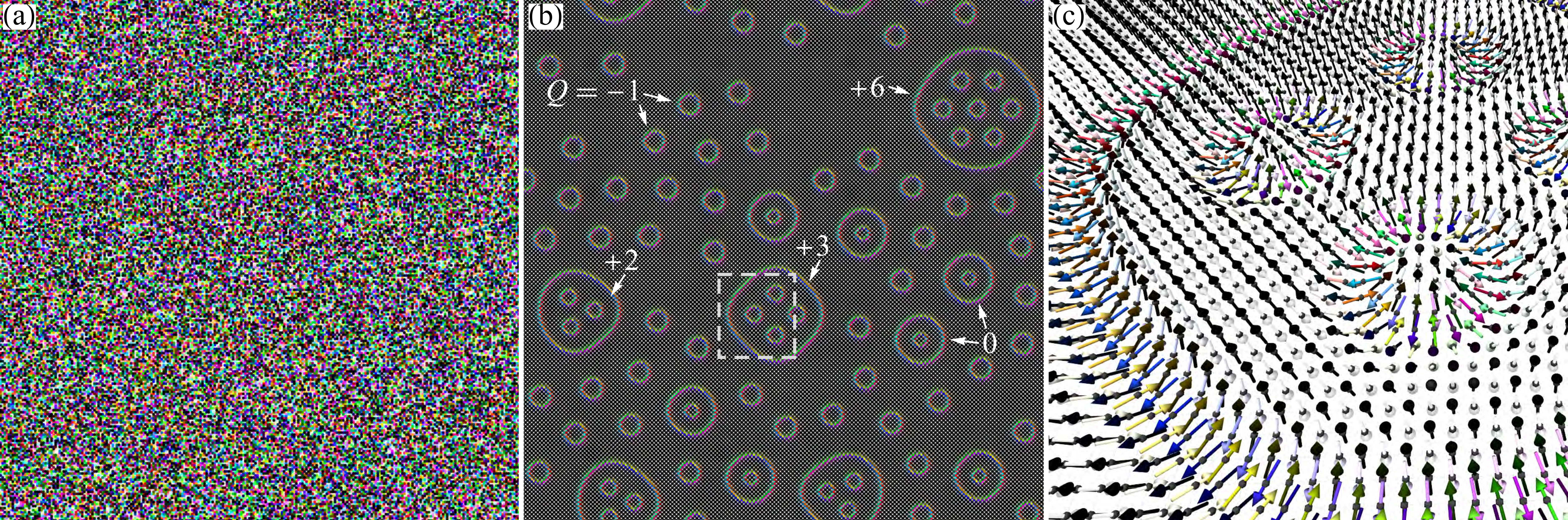}
\caption{\small
The emergence of antiferromagnetic skyrmions with different topological charges as a result of full energy minimization  starting from a random spins distribution.
(a) The initial entirely random spins distribution with zero net magnetization, (b) the spin configuration after complete minimization and (c) the perspective and zoomed view of the area marked in (b) as the dashed square.  
%
%
Calculations have been performed on the domain with $256\!\times\!256$ spins and periodic boundary conditions in the plane.
Texture with $Q\!=\!0$ (antiferromagnetic skyrmionium) has been recently discussed in~\cite{Bhukta}. 
} 
\label{sup_Fig_anti}
\end{figure}

The spin-orbit interaction in antiferromagnets plays a similar role in the stabilization mechanism of skyrmions as in ferromagnetic mediums.
Furthermore, the most realistic cases of two-sublattice chiral antiferromagnet can be described in the frame of the same effective model as for chiral ferromagnet~\cite{AntiferroMain, Bogdanov_PRB_2002}.

For the simulation of antiferromagnets, we used spin lattice Hamiltonian~(\ref{Ham_latt}) with $J\!=\!-1$, $D\!=\!0.4$, $K_u\!=\!0.22$ and $B_\mathrm{ext}\!=\!0$. 
In the corresponding phase diagram of a 2D chiral antiferromagnet, these parameters belong to the domain of confident stability of antiferromagnetic skyrmions~\cite{Bessarab_anti}.

For an antiferromagnet, in the absence of external magnetic field, the net magnetization reduces to zero. 
Therefore, it seems natural to use the configuration that meets this criterion as the initial state. 
Random spins distribution represent an inexhaustible set of initial states with a zero net magnetization.
It turned out that such simple initial guess with a regular probability leads to the appearance of skyrmions with various $Q$  after direct energy minimization, see Fig.~\ref{sup_Fig_anti} and \href{http://www.youtube.com/watch?v=xK_BSP8GX3c}{\underline{movie~4}}. 
The topological charge of antiferromagnetic skyrmion can be calculated for either of the two sublattices and taking into account its polarity (net magnetization of sublattice).
Because of the opposite polarities of the sublattices, the \textit{superimposed} or \textit{combined} winding number in this case always vanishes,  $-Q\!+\!Q\!=\!0$. 
An important consequence of such vanishing of winding number is the cancellation of so-called Magnus force -- the force acting on topological magnetic soliton interacting with the spin-polarized electric current~\cite{Zhang_anti,Barker_anti}.
Note, cancellation of Magnus force is expected for any antiferromagnetic skyrmions irrespective of topological charge $Q$.

The feature of topological charge for an antiferromagnet can be illustrated using Fig.~\ref{sup_Fig_anti} as follows.
Let us consider skyrmion with $Q\!=\!+3$ and three skyrmions with $Q\!=\!-1$ nearby. 
The total topological charge of such four textures is zero, which means that there is a way to merge these textures with further transformation into the ground state under preservation of the continuity in each of the sublattices.

\part*{\centering \normalsize Acknowledgments}
The authors thank Cyrill~B. Muratov, Egor Babaev, Stavros Komineas, Juba Bouaziz and Christof Melcher  for useful discussions of results. 
The work of F.\,N.\,R. was supported by the Swedish Research Council Grant No. 642-2013-7837, by G\"{o}ran Gustafsson Foundation for Research in Natural Sciences and Medicine  and by the ``Roland Gustafssons Stiftelse f\"{o}r teoretisk fysik''. 
The work of N.\,S.\,K. was supported by Deutsche Forschungsgemeinschaft (DFG) via SPP 2137 ``Skyrmionics'' Grant No. KI 2078/1-1.

\part*{\centering \normalsize References}

\renewcommand{\section}[2]{}

\end{document}